\documentclass[acmlarge]{acmart}

\makeatletter
\newcommand{\myconfshort}{\acmConference@shortname}
\newcommand{\myconffull}{\acmConference@name}
\newcommand{\myconfdate}{\acmConference@date}
\newcommand{\myconfloc}{\acmConference@venue}

\AtBeginDocument{
  \fancypagestyle{firstpagestyle}{
    \fancyhead{}%
    \fancyfoot[C]{}%
  }
  \fancyhf{}
  \fancyhead[LO]{\@headfootfont\shorttitle}%
  \fancyhead[RE]{\@headfootfont\@shortauthors}%
  \fancyhead[LE]{\@headfootfont\footnotesize \myconfshort, \myconfdate, \myconfloc}%
  \fancyhead[RO]{\@headfootfont\footnotesize \myconfshort, \myconfdate, \myconfloc}%
  \fancyfoot[C]{}%
}
\makeatother

\acmBooktitle{\conffull\@ (\confshort), \confdate, \confloc}

\usepackage{array}
\usepackage{enumitem}
\usepackage[title]{appendix} 
\usepackage{array}
\usepackage{enumitem}
\usepackage[title]{appendix}
\usepackage{booktabs}
\usepackage{makecell, booktabs, tabularx}
\usepackage{tabularx, booktabs, array}
\usepackage{booktabs}
\usepackage{tabularx}
\usepackage{graphicx} 
\usepackage{array}
\usepackage{soul}
\date{}

\copyrightyear{2026}
\acmYear{2026}
\setcopyright{cc}
\setcctype{by}
\acmConference[FAccT '26]{The 2026 ACM Conference on Fairness, Accountability, and Transparency}{June 25--28, 2026}{Montreal, QC, Canada}
\acmBooktitle{The 2026 ACM Conference on Fairness, Accountability, and Transparency (FAccT '26), June 25--28, 2026, Montreal, QC, Canada}
\acmDOI{10.1145/3805689.3812395}
\acmISBN{979-8-4007-2596-8/2026/06}

\begin{document}

\title{Guardrails versus Gatekeepers: Understanding Product Managers' Ethical Decision-Making in Generative AI}

\author{Genevieve Smith}
\authornote{These authors share lead authorship.}
\email{genevieve.smith@berkeley.edu}
\affiliation{%
  \institution{University of California, Berkeley}
  \city{Berkeley}
  \state{CA}
  \country{USA}
}

\author{Natalia Luka}
\email{nataliyan@berkeley.edu}
\authornotemark[1]
\affiliation{%
  \institution{University of California, Berkeley}
  \city{Berkeley}
  \state{CA}
  \country{USA}
}

\author{Merrick Osborne}
\affiliation{%
  \institution{Cornell University}
  \city{Ithaca}
  \state{NY}
  \country{USA}
}

\author{Brian Lattimore}
\affiliation{%
  \institution{Stanford University}
  \city{Palo Alto}
  \state{CA}
  \country{USA}
}

\author{Jessica Newman}
\affiliation{%
  \institution{University of California, Berkeley}
  \city{Berkeley}
  \state{CA}
  \country{USA}
}

\author{Brent Mittelstadt}
\affiliation{%
  \institution{University of Oxford}
  \city{Oxford}
  \country{UK}
}

\author{Brandie Nonnecke}
\affiliation{%
  \institution{University of California, Berkeley}
  \city{Berkeley}
  \state{CA}
  \country{USA}
}

\renewcommand{\shortauthors}{Smith et al.}
\acmArticleType{Research}
\acmContributions{}
\begin{abstract}
\section*{Abstract}
What is the role of product managers (PMs) in the responsible use of generative AI (genAI) in products and everyday work---and what enables or constrains their ability to take action? Past literature has examined the ways in which organizational policies can become decoupled from practices when incentives for responsible action are misaligned or impeded by profit motives. While the role of engineers and professional ethicists in the context of AI has been examined in detail, the role of product managers---who are frequently portrayed as “gatekeepers” or critical decision-makers in product teams---remains unclear, particularly regarding genAI. In this paper, we examine what organizational conditions promote responsible use of genAI by product managers by drawing on twenty-five interviews and a global survey of over three hundred respondents in product management-related roles. First, we find that uncertainty around responsible AI and a sense of diffused responsibility constrain ethical action, while leadership commitment and organizational principles enable ethical action---making some responsible practices up to fourteen times more likely. Further, our study finds two sets of actions product managers take to “recouple” ethical commitments and practices. The first set includes \textit{low-resource, individual actions} product managers can implement without explicit organizational incentives (e.g., individual or team-wide reviews and safeguarding standards around data privacy). The second set includes \textit{high-resource, collective actions} that require organizational incentives (e.g., conducting audits or delaying shipment of products). In contrast to existing findings in the literature, product managers view themselves as "guardrails" rather than “gatekeepers” when it comes to making ethics-related decisions for genAI use. Our research suggests that recoupling ethical policies and practices at the level of product teams requires institutional buy-in and higher level leadership commitment. 
Nevertheless, we show that individual actors are able to exhibit agency through some meaningful, low resource actions, even in the absence of organizational incentive structures even in the absence of organizational incentives, though this alone is insufficient to operationalize responsible AI at scale. 
\footnote{This is the author's version of the work. This preprint is posted here for your personal use. The definitive Version of Record is published in the 2026 ACM Conference on Fairness, Accountability, and Transparency (FAccT '26), June 25--28, 2026, Montreal, QC, Canada.}
\end{abstract}

\maketitle

\section{Introduction}
As the adoption of AI has grown, so has the idea of “responsible AI” \cite{jobin2019artificial, mittelstadt2019principles}. Responsible AI (sometimes used synonymously with “ethical AI”)  refers to the development, deployment, and governance of AI technologies to align with human values and societal norms \cite{deshpande2022responsible, birhane2022power}. Past studies have examined the phenomenon of “ethics washing” and the ways in which organizational policies can become decoupled from practices when incentives for responsible action are misaligned or impeded by profit motives \cite{greene2019better, metcalf2019owning, attard2023ethics, neff2020bad}. Practitioners, ranging from software engineers to professional ethicists, have struggled with a lack of structural power when questioning whether to uphold their ethics and values in the workplace \cite{rakova2021where, nedzhvetskaya2022role, widder2023about, ali2023walking}. 

Product managers are the dominant form of mid-level manager within product-focused organizations, particularly within the software industry and software-focused divisions of companies \cite{maglyas2013what}. Existing studies have noted that product managers play a privileged role in the power structures of most tech companies, serving as gatekeepers, controlling most day-to-day decision-making in the product development life cycle.\cite{rakova2021where, ali2023walking, widder2023about} Ali, et al. (2023) found that professional ethicists spent a significant amount of their time “negotiat[ing] with product managers for resources, theoriz[ing] ethics-oriented interventions as functionally superior or “pragmatically legitimate” by linking them to product quality, and advocat[ing] for early integration into the product development cycle”. Widder, et al. (2023) found that, among software engineers, product managers were viewed as “decision-makers” that had to first be convinced in order for certain ethical actions and interventions to be carried out \cite{widder2023about}. While product managers are viewed as powerful actors by employees in other roles, we know little about how they conceptualize their own agency and what factors drive them to take actions prioritizing ethics. This is particularly the case for generative AI (genAI), a form of AI that generates new outputs from existing data, including text, code, images, and patterns of behavior.\cite{kenthapadi2023generative} What is the role of product managers in integrating, promoting, and challenging ethical use of genAI technologies in the workplace? What constrains or enables product managers in the types of actions they choose to take towards ethical use of genAI? 

In this paper, we examine the persistent research gap in understanding how responsibility is incorporated in organizational decision-making among product managers, and the conditions that constrain or support responsible development and use of generative AI (genAI). Drawing from our empirical data, we find two categories of actions product managers take to “recouple” ethical commitments and practices regarding genAI. The first category includes low resource, individual actions product managers can take without explicit organizational incentives, which include individual or team-wide reviews and safeguarding standards around data privacy. The second category includes  high resource, collective actions that require organizational incentives, which include conducting audits, fairness/bias testing, or adversarial testing, such as red teaming. 

Our research highlights both the possibilities and limitations of such actions, suggesting that even in environments that are not conducive to higher-resource actions, there remain ways for employees to exhibit agency in promoting ethical genAI use. At the same time, many actions are realistically off-limits without broader organizational buy-in. We counter the existing conception of PMs as \textit{gatekeepers }within the AI ethics and governance literature (individuals who determine critical decisions in the product development cycle) and suggest that the more realistic view is of product managers as \textit{guardrails }(individuals who uphold organizational standards and principles set by higher-level leadership rather than exercising independent ethical authority). Rather than calling for reduced accountability for PMs, these findings call for greater organizational support to create the structures and conditions that make ethical robust action at the product team level genuinely possible.

\section{Background}
\label{sec:related-work}
Research on AI ethics and governance in organizations has documented the challenges of translating high-level ethical principles and policies into everyday organizational practice. This background section outlines prior work on organizational decoupling and recoupling in AI ethics, and situates product managers as critical, yet understudied, actors navigating gaps between ethical principles and practices. 

\subsection{Convergence on principle, decoupling in practice}
Principles for AI ethics and governance---across institutions and in different country contexts---show a surprisingly high amount of convergence, with alignment on concepts including fairness, safety, accountability, privacy, and transparency \cite{jobin2019artificial, hagendorff2020ethics, reuelrai}. While these principles form a useful background that allows for dialogue across different spheres and sectors, challenges remain in the operationalization and implementation of responsible AI principles within organizations. \cite{lancaster2024accountability}. Principles can be thought of as precursors to policies, which prescribe concrete actions for actors within organizations. Decoupling occurs in organizations when there is a gap between what is stated as a principle or policy and what is carried out in the day-to-day practices of an organization) \cite{rakova2021where, nedzhvetskaya2022role, ali2023walking, rustergaps, dotan2025decoupling}. 

This gap has been criticized as “ethics washing”, in which organizations promote the image of responsible or ethical action for reputational benefit \cite{greene2019better, metcalf2019owning, phan2021economies,seele2022greenwashing, neff2020bad}. Ethics washing is associated with low-resource, low-risk activities that have minimal impact on an organization's day-to-day operations, products, or services. \cite{attard2023ethics} Certain practices lend themselves more easily to ethics washing than others, for example, creating high-level ethical principles. Principles have been criticized in the literature for being easy to construct but challenging to uphold, often because they remain too abstract and are not tied to specific practices or goals \cite{morley2023operationalising}. Internal audits, by contrast, have been noted as being one of the more challenging ethical practices to implement and there is evidence that they are more rarely adopted by organizations as a result \cite{mokander2021auditing, costanzachock2022auditors}. However, it's critical to note that what distinguishes ethics washing from genuine ethical action is the gap that occurs when principles and policies are not converted to regular practices, not the practice in and of itself. \cite{schultz2025ethicswashing} 

This gap between principles and practices can be the result of a range of structural and cultural barriers within organizations. At a high-level, organizational cultures prioritizing “speed to market” and profitability can relegate ethics as a peripheral concern \cite{west2019discriminating}. Within such structures, a moral hazard can arise when actors believe others are accountable for ethical issues or when their roles lack incentives to prioritize ethical considerations \cite{holmstrom1979moral}. Decoupling, however, can also be the result of bottom-up as opposed to top-down decisions on ethics, where individual actors operate in environments of high uncertainty, misaligned incentives, or poorly defined and improperly resourced roles \cite{rakova2021where, nedzhvetskaya2022role, kuehnert}. These actors often face organizational inertia and unclear mandates when AI ethics are not integrated into other incentive structures such as performance metrics \cite{ali2023walking, zhou2023ai}. Finally, a diffusion of accountability can occur when accountability is fragmented across teams and responsibility is perceived as shared without clearly defined roles. \cite{darley1968bystander, grote2024taming} 

\subsection{Organizational conditions enabling recoupling in responsible AI}

In organizational sociology, the re-integration of ethical policies in day-to-day practices is known as “recoupling”, a dynamic process through which commitments become meaningfully enacted in routine work \cite{hallett2010myth}. In broader organizational and sociological literature, recoupling typically requires one of two conditions: (1) external pressures that hold actors accountable to outside stakeholders, such as regulation, audits, or public scrutiny \cite{espeland1998struggle, hallett2010myth, turco2012difficult} or (2) commitment from higher management \cite{weaver1999integrated}. One or both should ideally be operationalized as internal incentives in order to ensure compliance. 

In the context of AI ethics and governance, these conditions are often weak or missing. At the time of writing, regulation of AI is limited, though growing, and lack of transparency into foundation models, tools, and systems prevents audits and public scrutiny. \cite{wan20252025foundationmodeltransparency, shen2025disclosure} Leadership support, when present, is rarely instrumentalized into concrete incentive structures and performance metrics. \cite{morley2023operationalising, ali2023walking}. Examples to the contrary, of course, exist, such as Sloane and Zakrzewski's 2022 study of ethical AI within German start-ups. The authors find that when ethical practices were successfully implemented, it was because they were conceived of as both an individual and collective responsibility, drawing upon group decision-making processes that are an established part of German corporate governance culture. \cite{sloane2022germanai} In this case, leadership commitment is integrated into governance structures but, as the author themselves note, we should not assume these structures exist in other organizations or geographies.

However, even as the field has recognized the challenges to ethical implementation, it has attempted to design interventions to aid recoupling. These include accountability frameworks that assign responsibility for ethical risks and ethics-by-design approaches that translate values into design requirements \cite{morley2020whattohow, morley2023operationalising, brey2024ethics, madanchian2025ethical, papagiannidis2025responsible}. These 
interventions draw heavily from the human-computer interaction 
(HCI) field, which has produced a range of practical tools and methods aimed at operationalizing ethical principles at the level of everyday product work. \cite{suchman1995makingwork, ackerman2000, friedman2017, fox2020workercentered, shneiderman2020} These interventions are not sufficient on their own, however, to close the gap between ethical principles and practice. Absent external accountability pressures or leadership commitment, they are unlikely to drive meaningful recoupling. \cite{ali2023walking, rakova2021where}. They can, however, create 
helpful processes for leaders to articulate clearer and more 
concrete expectations around responsible AI practice.

\subsection{The critical role of product teams and managers}
In the late 20th century, product managers emerged as a dominant form of management in product-focused organizations, particularly within the software industry and software-focused divisions of companies. While most mid-level management structures developed to enforce strict hierarchies in manufacturing processes, product management developed as a means to oversee the cross-functional, relatively egalitarian teams more common to software development. \cite{mcdaniel1980product, maglyas2013what} This led to the development of a management style that was more collaborative and project-oriented, allowing for greater discourse across functions, as well as work cultures that lent themselves to more outspoken ethical expression \cite{tarnoff2020making, tan2025unlikely}). 

Product managers have been viewed as central figures in the day-to-day decision-making processes at most companies specializing in software development, which includes the development of genAI models and systems. Their structural position is distinctive: sitting between higher-level leadership and cross-functional execution 
teams, PMs control priorities, allocate resources, and determine whether and how ethical considerations are built into AI products \cite{rakova2021where, ali2023walking, 
widder2023about}. This intermediary positioning has led other members of product teams, including software engineers, designers, and professional ethicists, to treat PMs as the critical decision-makers when ethical concerns arise. In Ali et al's (2023) account, professional ethicists within organizations may advocate for responsible practices, but ultimate decision-making authority is understood to rest with the product manager.

Across this body of work, product managers are consistently portrayed as \textit{gatekeepers}: actors with the authority and organizational leverage to determine whether a product is developed or shipped, and therefore whether ethically fraught decisions are made or deferred. This framing positions the PM as agents who exercises independent ethical judgment at key junctures in the product development lifecycle, a chokepoint through which responsible AI commitments can either be enforced or abandoned. Yet this characterization raises an important and underexplored question. The gatekeeper framing reflects how PMs are perceived by others on their teams; it does not necessarily capture how PMs understand their own role and the scope of their agency. If product managers operate under conditions of uncertainty, misaligned incentives, or diffused responsibility—as the broader literature on organizational decoupling suggests is common in AI ethics contexts—the degree to which they can exercise independent ethical authority may be substantially more constrained than the gatekeeper model implies. The distinction matters: a gatekeeper who lacks the organizational conditions to act as one is not simply a weak gatekeeper, but a different kind of actor altogether. How PMs themselves navigate this gap, and what it means for the ethical governance of genAI, is the question this paper sets out to address.

\section{Methodology}
\label{sec:methodology}
\subsection{Analytical Approach}
We employ a mixed methods analytical framework drawing on interview and survey data. Qualitative interview data are analyzed through thematic analysis coded with NVivo software using an abductive approach \cite{tavory2012theory}. Quantitative survey data are analyzed using descriptive and inferential statistical techniques, including logistic regression, to identify patterns between organizational conditions and adoption of responsible AI practices. The research protocol was reviewed and approved by the ethics review committee at the lead author's university prior to data collection. All participants, for both the survey and interviews, reviewed and signed informed consent and confidentiality agreements. 
We also preregistered the survey with the Open Science Framework to ensure transparency in our methods.\footnote{We have anonymously registered the study with the Open Science Foundation via the following link: \url{https://osf.io/xfwka/overview?view_only=bc1181f1ced94dc7aa479ed74d80e6a4}}

\subsection{Methods and Data}
\textit{Interviews.} We conducted 25 semi-structured interviews with product managers at companies across a range of industries, predominantly tech. Since the purpose of the study is to understand organizational decision-making by product managers, the key eligibility requirement for participants is that they are in a product management position (or were within the last six months) and are using genAI in their everyday work and/or integrating genAI into new products. We adopted a broad definition of "using genAI technologies" that included the use of any genAI tools, such as chatbots and AI assistants or agents.

Other inclusion criteria included: employment at the organization for at least 3 months, being over 18 years of age, providing informed consent, and fluency in English. There were no inclusion or exclusion criteria based directly on health status, gender, race, or ethnicity. Participants were all located in the United States and interviews were conducted via Zoom, lasting between 45 minutes and one hour and were not compensated. The call for interview participants was distributed by a large global technology company headquartered in California to employees of the company as well as partners with the assurance that data collected would not be seen by the company. Further recruitment was conducted through purposive sampling, via a call for interview participants distributed via email, and snowball sampling, via recommendations from prior interviewees. 

The full corpus of material analyzed included interview transcripts and post-interview analytic memos. Interviews were transcribed and edited to retain substantive content while omitting filler words and non-verbal utterances that did not add additional meaning.
All identifying information was removed from materials following data collection. Data were stored on a password-protected cloud storage system.

We conducted a reflexive thematic analysis, developing, analyzing, and interpreting patterns across the data to draw out themes [4]. Following an initial generation of codes informed by our theoretical basis and literature review, we reviewed themes from the codes in NVivo to identify patterns and core elements to then refine the codebook and writing. We utilized abductive reasoning with a grounded theory approach to allow important themes to emerge \cite{tavory2012theory}. 

All those interviewed had product management titles, with varying degrees of seniority. Of those interviewed, 40\% were product managers, 36\% were at senior levels of product management, and 24\% were director levels of product management. 60\% of interviewees were male and 40\% female. The vast majority of interviewees are in the tech industry (80\%), with a sampling in other industries: media (8\%), finance (4\%), real estate (4\%), and education (4\%). All interview participants are based in the United States, with the majority in California (36\%) followed by New York (24\%).

\textit{Survey.} The call for participants was posted on Prolific outlining the inclusion criteria, which was the same as the interview criteria. We used Prolific pre-screening to include eligible countries for the survey and selected the employment domain as “product management”. This way only people who self-identified as employed in “product management” received the ad. Once eligible participants self-selected the task, they were provided with a link to the survey in Qualtrics. While it is possible that interview participants may have also completed the survey, this is unlikely for several reasons: the interview sample was small (n=25), the studies were conducted sequentially with interviews preceding the survey, and the two samples were recruited through entirely different channels. The survey included a consent form and confirmation of meeting the inclusion criteria. No identifiable information was collected from subjects and IP collection was turned off to protect confidentiality. Participants received US\$4 upon completion of the survey. This amount was based on a US\$16/hour  wage at an estimated survey completion time of 15 minutes.

We collected basic demographic information about survey participants (n=303). 49.5\% of respondents identify as male, 48.84\% as female, and 1.32\%  as nonbinary. The majority of respondents identify as White (61.72\%), followed by Black (15.51\%), South Asian (8.58\%), East Asian (5.61\%), multiracial (3.63\%), and American Indian (<1\%). Most are based in the United States (53\%), followed by the United Kingdom (26\%), and Canada (10\%). The remaining respondents drew from other countries in our inclusion criteria including: Germany, India, Denmark, Finland, France, Scotland, Ireland, Sweden and Spain. Most are aged 25 to 34 (38.94\%, followed by 35 to 44 (29.7\%), 45 to 54 (15.1\%), 18 to 24 (8.25\%) and over 65 (1.65\%). While the task was shown to those in Prolific who self-identified in employment areas of “product management”, we enquired “what best describes your role” and the majority selected product manager (49.17\%), followed by product marketing (23.76\%), product designer (8.25\%), product engineer (7.6\%), and other (10.9\%). Respondents represent a range of industries, particularly IT (19.8\%), manufacturing (15.8\%), retail (12.21\%), finance (10.23\%), arts (6.6\%), healthcare (5.61\%), transportation (4.6\%), education (4.3\%), and other (8.6\%). 

In our logistic regression, we regressed the likelihood of various responsible AI practices being adopted by a respondent in our survey sample (dependent variable) against the set of organizational conditions that the same respondent reported at their workplace (independent variables). We chose to run a logistic binary regression given the binary nature of outcomes (0 = no adoption, 1 = adoption of a practice) and the ability to control for variables, which include firm size and industry fixed effects. There are obvious challenges to working with binary variables because they can obscure the high degrees of nuance and contextual complexity that drive organizations to adopt certain ethical practices and not others. For this reason, we situate our regression results after our analysis of interview results, which address this kind of complexity. Our coefficients are represented as exponentiated log-odds ratios. If the coefficient is greater than 1, it indicates that the presence of a certain organizational condition increases the likelihood of a practice being adopted. If the coefficient is between 0 and 1, it indicates that the presence of a certain organizational condition decreases the likelihood of a practice being adopted. The responsible AI practices in the survey included a list of those frequently cited across five common responsible AI topics and principles (i.e. fairness, security and safety, data privacy, transparency, and accountability). 

\section{Findings}
\label{sec:findings}
We first present findings on three key barriers for product managers to use genAI responsibly, before turning to the enablers. We then delve more specifically into what kinds of actions are constrained or enabled under different organizational conditions. 

\subsection{Barriers to responsibility: Uncertainty}

One of the key findings across both interviews and surveys was a pervasive sense of uncertainty regarding what responsibility means and looks like in using genAI. Of the survey respondents, 76.57\% said lack of clarity on what ‘responsibility’ looks like was a challenge they or their team faced in regard to using genAI responsibly (see Table 1). Our qualitative data reveals that this uncertainty is tied to three factors which we further explore in this section: (1) immaturity in the industry around genAI adoption, (2) lack of transparency regarding the foundation models, and (3) lack of clarity about what responsibility means from an organizational perspective.

\begin{table}[ht]
\centering
\caption{Participant experiences of different challenges in using genAI responsibly}
\label{tab:genai-challenges}
\begin{tabular}{@{}lr@{}}
\toprule
\textbf{Challenge Category} & \textbf{\% of Respondents} \\ \midrule
Lack of clarity on what 'responsibility' looks like & 76.57\% \\
Lack of resources or tools & 46.20\% \\
Lack of training & 39.27\% \\
Lack of incentives & 36.96\% \\
None & 27.39\% \\
Lack of trust (personally) in the genAI tool/model itself & 26.73\% \\
Lack of understanding whether / why it may be valuable & 21.12\% \\
Lack of support & 17.49\% \\
Lack of clarity about expectations or relevance to my role & 17.16\% \\ \bottomrule
\addlinespace[1ex]
\multicolumn{2}{@{}l}{\footnotesize \textit{Note: Respondents could check all that apply.}} \\
\end{tabular}
\end{table}

In our interview, there was a noticeable lack of aligned definitions and concrete guidance to using genAI responsibly, reflecting the immaturity of the industry (factor 1). One interviewee, Jae\footnote{All names used in this paper are pseudonyms to ensure we preserve the anonymity of our participants.}, noted, “Responsible AI is a buzzword, right? But it's a buzzword because it doesn't actually mean what people think it means. And it's really elusive and vague and abstract.” This captures the sense of uncertainty around the concept of “responsible” AI both at the individual level, as well as more broadly in industry where responsibility lacks aligned, concrete guidance despite the popularity of the term. 

Uncertainty also stems from opaqueness that permeates the AI space. In particular, uncertainty on responsible use is tied to the lack of transparency in prominent foundation models (factor 2), whose developers and/or owners of such models limit public sharing of training data, model architectures, and decision-making processes. Without having access to this key information about the model it is difficult to know what types of responsibility risks exist in the model that may be adopted in its use. Jack captured this well: 
\begin{quote}
“There isn't enough transparency today in [genAI] technologies… Almost none of the open models are actually open source to the point where I can look at them and know exactly what's inside of them and what I'm using, I can't tell what biases went into tuning those machines, I don't know what values were applied in order to add safety, guardrails or behavioral guardrails… There's an information asymmetry that has to be addressed. We need more transparency from the makers of this technology… We're trying to bring various ethical and moral frameworks and legal frameworks to a thing we don't have enough data about.” 
\end{quote}

Uncertainty is also tied to a third factor: lack of clarity from an organizational perspective. This was particularly expressed by product managers integrating genAI into new products or features. Jae added, “A lot of people don’t know what to test. They don’t know why they need to test [genAI]. They don’t even know how to conceptualize risk… Okay, you’re shipping this thing with AI, how is it going to harm people?” This quote captures the broader uncertainty around measuring harm, linked to immaturity in the industry, and also the lack of clear guidance from one’s organization resulting in individuals taking on their own risks. Not surprisingly, following lack of clarity, respondents most frequently noted a lack of resources (46\%) and lack of training (39\%), suggesting that some of the challenges to using genAI responsibly stem from a general unfamiliarity with the technology coupled with the inability of organizations to provide sufficient resources to bridge this gap.

\subsection{Barriers to responsibility: Diffused responsibility}

Interestingly, some participants reflected in the interviews that their own lack of action on responsible AI was tied to the belief that other teams are handling it. 
A portion of product managers, usually at larger and more specialized organizations, believed other teams were tasked with making sure AI is used responsibly within the organizations and its products. Because of their position at the cross-section of multiple teams, our findings suggest that product managers are particularly susceptible to this mindset compared to other roles. Without clear organizational guidance at the product level, there can be a sense that responsibility is taken care of by other teams. Peter noted, “There’s just so many different teams, so much bureaucracy before something goes to market... So I guess I’m just trusting my colleagues have already approved this with the mark of ‘yes, we’re being responsible by using this tool.’ ” 
This is notable given prior literature noting that product managers are often viewed as key decision-makers in how AI is integrated into products. Yet these findings suggest that even those with significant influence over product direction do not necessarily see responsible AI as within their purview, assuming it is handled elsewhere while often lacking the organizational clarity to act on it themselves.

\subsection{Barriers to responsibility: Lack of incentives}

Another prominent constraint to the responsible use of genAI, identified in both interviews and the survey, was a lack of incentives. Specifically, required action items were unclear or absent, and responsibility for ethical oversight was not meaningfully aligned with rewards. Sarah put it simply, “There’s no clear action item sometimes when we’re talking about responsibility and ethics.” Instead, incentives for product managers are clearly tied to efficiency and product delivery. Jae captured, 
\begin{quote}
“It's really scary if everyone's just shipping AI stuff all the time. Right. It's not auditable, it's not tracked… I want to get promoted, but I can't because I can't ship this thing and promotions and rewards are very tied to shipping stuff. And so if you stop people from shipping stuff and then you don't tell them how to move forward with that, then it's just  a disaster for the company.” 
\end{quote}
Interviewees frequently observed that incentive structures can conflict with higher-level principles and values underlying responsible AI use. They described a difficult trade-off between short-term, measurable outcomes (e.g., shipping products and delivering company growth) and long-term, more abstract goals, such as upholding ethical standards and mitigating potential future harms. Since product managers bear the responsibility for preparing products to "ship", i.e. reach customers, they may be uniquely positioned to comment on this challenge compared to other individuals in other roles.

\subsection{Adoption of responsible AI practices }
Despite working in an environment of high uncertainty, diffused responsibility, and lack of incentives, approximately 80\% of survey respondents reported adopting at least one responsible AI practice. We found that considering data privacy implications and safeguarding data privacy was the most commonly adopted practice with slightly less than half (46.9\%) of the respondents reporting that they have taken this action in their work. This was followed by taking ethical/responsible AI trainings (42.6\%) and asking about AI data or models to understand their limitations and potential issues (38.3\%). The least commonly reported actions were conducting audits of genAI tools (25.4\%) and conducting adversarial testing or red teaming (14.2\%). 

\begin{table}[ht]
\centering
\caption{Participant adoption of different practices using GenAI responsibly}
\label{tab:genai-practices}
\begin{tabular}{@{}p{0.75\columnwidth}r@{}}
\toprule
\textbf{Responsible AI Practice} & \textbf{\% of Respondents} \\ \midrule
Consider data privacy implications \& take actions to protect data privacy & 46.9\% \\
Take ethical / responsible AI trainings & 42.6\% \\
Ask about the data or model to understand potential limitations/issues & 38.3\% \\
Build transparency approaches (e.g., use documentation for datasets/models) & 30.7\% \\
Use explainability methods (e.g., to enable team members to probe the model) & 30.4\% \\
Conduct fairness / bias testing & 30.0\% \\
Conduct audits of the GenAI tools & 25.4\% \\
No actions taken & 19.8\% \\
Conduct adversarial testing or red teaming & 14.2\% \\ \bottomrule
\addlinespace[1ex]
\multicolumn{2}{@{}l}{\footnotesize \textit{Note: N = 303. Respondents could check all that apply.}} \\
\end{tabular}
\end{table}

Individual product managers discussed the conditions that led them to adopt some practices and not others. We found that the actions that were most commonly adopted were typically low resource, individual actions in-line with incentives or low cost in terms of time and effort. In our interview findings, these included practices such as conducting gutchecks of analyses or outputs from LLMs and safeguarding private data from proprietary LLMs. As Jorge described these practices: “you need to be aware of…the data that you're giving to the model… and the training data that you're using and at the same time be aware of misinformation and misuse, so basically the output of the model, if that makes sense or not”. Compared to larger interventions, these were lower risk since they tended not to slow down the product development cycle and did not require buy-in from higher level managers. 

In some instances, organizations subsequently formalized these practices, such as by adopting evaluation tooling that teams are expected to use or instituting structured reviews with external teams. Michael described a tool within their organization: 
\begin{quote}
“every single bit of data gets tracked in that tool, and that's more like the PM tool, where you're essentially going through and you're making the data requests, you're adding any terms or licenses or this is how the data was collected…and then once it gets approved, that number goes into the record for the product…if there's any security review or compliance comes in or accounting has any questions it's like, here's the number.” 
\end{quote}
In addition to being embedded in organizational work processes, this example incentivized use by making records subject to potential audit by compliance teams or regulators.

Despite being perceived as lower risk relative to larger interventions, these actions nevertheless did help address accountability gaps and provide ways for product managers to exhibit control over responsible development and use of genAI on their teams. Interviewees suggested ways these practices could be incorporated in their jobs---such as clear documentation and articulation of company policies. They noted there was also the potential to address broader, industry-wide uncertainty through the creation of shared standards. As Aditya described, “we have an opportunity as an industry to say, ‘OK, these are the standardized benchmarking frameworks, datasets, metrics or content safety evaluation, content quality, etc.’... There is an opportunity here to bring this together in the best interest of customers or people in general”.

\subsection{Enablers to responsibility: Leadership commitment and incentives}

While survey findings allowed us to understand the context in which most product managers were working, interview findings illustrated how they made sense of this context and the ways in which they were able to operationalize responsibility in spite of uncertainty, diffused responsibility, and a lack of incentives. Important conditions for operationalizing responsibility include having principles tied to company values and leadership commitment, as well as incentives with tools and frameworks. 

Both interview and survey results supported the importance of leadership commitment. As Arjun said, “Leaders set the tone… that expectation that we're not gonna be cowboys about this, we're gonna think about this the best we can. Then it frees everyone else in the organization to say the thing they're worried about.” Clear leadership committed to responsibility signals to the rest of the organization that it is a priority and supports a culture of collective responsibility.  While not sufficient in and of themselves, principles play a key role when partnered with leadership commitment, particularly as the industry continues to move quickly while much uncertainty remains. Aditya said, 
\begin{quote}
“How do you balance speed versus responsible AI? I don't know what the best solution is…but a solution that works better than others is having the top down trusted AI principles and guidance. That becomes a mandate for bottom up and deployment life cycles. If leadership is not putting their foot down to say this is important and critical, it's going to be hard for the builders to enforce that… Every PM in this field is thinking about that trade off. How fast am I moving and how can I balance that with responsibility and doing this in a trustworthy way?” 
\end{quote}

While pressures remain to move quickly and traditional incentives focus on shipping products, principles create opportunities for product managers to bring in other priorities and validate ethical actions. Interviewees, however, were also quick to note a gap, or a decoupling, between a company’s principles and their practices. As Jill shared, “I think if you look at any given company’s mission, statement, and other highfalutin documentation and then look at what they're actually doing. Again, there is a void in between the two things”. Interviewees frequently commented that it was far easier to establish principles than to define them in real life and put them into practice, especially when there was a lack of accompanying metrics and incentives. As Jack noted, 
\begin{quote}
“Management has to say, 'this is important to me'. They have to say, 'this is how you'll be measured as an employee', because otherwise everyone's looking at their OKRs, their KPIs and it's all about growth, growth, growth and speed, speed, speed… At the end of the day it's going to be company leadership telling employees that they value this. And I think once they do that, the rest will fall into place.”
\end{quote}

In Table 3, we run a series of logistic regressions with each individual practice as the dependent variable and the set of organizational conditions (leadership commitment, principles, policies, incentives) as the independent variables to examine how different organizational conditions impact the adoption of responsible AI practices. The coefficients are shown as exponentiated log-odds ratios with p-values for significance. 

We find that leadership commitment, principles, and policies are significantly and positively associated with the adoption of all responsible AI practices we consider. This generally aligns with findings in our interviews that stressed the importance of all of these conditions. In particular, leadership commitment increases the likelihood of each responsible AI action between 1.94 and 4.42 times relative to no beneficial conditions (our baseline), illustrating how significant managerial buy-in is for responsible AI practice adoption. Similarly, having responsible AI principles increases the likelihood of each action by between 1.83 (data privacy) and 7.06 times (responsible AI training) relative to our baseline. Having responsible genAI policies increases the likelihood of each action by between 2.20 (genAI audits) and 14.44 times (adversarial testing) relative to our baseline. 

On the other hand, incentives were associated with taking some practices, but failed to reach statistical significance for others. They are tied to all practices except two: considering data privacy implications and building or using transparency approaches. These findings aligned with our interview results, suggesting that, absent organizational incentives, individual product managers will only be able to adopt certain actions, like the above. 

\begin{table}[ht!]
\centering
\scriptsize 
\caption{Exponentiated Log-Odds Ratio of Responsible AI Practices}
\label{tab:log_odds_equal_width}

\newcolumntype{C}{>{\centering\arraybackslash}X}

\begin{tabularx}{\textwidth}{@{}l|CCCCC|CCC@{}}
\toprule
 & \multicolumn{5}{c|}{\textbf{High-Resource, Collective Actions}} & \multicolumn{3}{c}{\textbf{Low-Resource, Individual Actions}} \\
\cmidrule(lr){2-6} \cmidrule(lr){7-9}
\textbf{Organizational Conditions} & 
\makecell{Audit\\ Gen AI} & 
\makecell{Bias\\ Testing} & 
\makecell{Adversarial\\ Testing} & 
\makecell{XAI\\ Methods} & 
\makecell{RAI\\ Training} & 
\makecell{Data\\ Privacy} & 
\makecell{Query\\ Limits} & 
\makecell{Transp.\\ Docs} \\ \midrule

Leadership commitment & 4.05*** & 2.76*** & 2.74** & 2.52*** & 4.42*** & 1.94** & 2.59*** & 3.80*** \\
RAI principles        & 2.82*** & 4.01*** & 4.16*** & 2.75*** & 7.06*** & 1.83* & 2.73*** & 4.14*** \\
Gen AI policy         & 2.20** & 3.36*** & 14.44***& 2.65*** & 6.26*** & 2.95*** & 2.36** & 3.41*** \\
Clear incentives      & 4.14*** & 2.05* & 2.72* & 1.87* & 2.83*** & 1.55    & 2.60** & 1.55    \\
None of the above     & 0.33** & 0.31** & 0.15* & 0.27*** & 0.06*** & 0.75    & 0.48* & 0.19*** \\ \bottomrule

\addlinespace[1ex]
\multicolumn{9}{@{}p{\textwidth}}{\tiny \textit{Notes: N = 303. All models include controls for firm size and industry fixed effects. Coefficients are exponentiated log-odds ratios.* p $\le$ 0.05, ** p $\le$ 0.01, *** p $\le$ 0.001.}} \\
\end{tabularx}
\end{table}

For the purpose of analysis, we divide our eight practices into two categories: high resource, collective actions versus low resource, individual actions. \textit{High resource, collective actions} typically involve teams of people with resources (training and tools) and include conducting audits, conducting fairness/bias testing, conducting adversarial testing or red teaming. Explainability methods and ethical/responsible AI trainings are also included in this category because, although they do not involve teams, they still require specialized training typically provided or contracted through the organization. High resource, collective actions, such as conducting audits, had the strongest positive association with organizational incentives. These findings highlight the importance of conditions such as leadership commitment, principles, and policies and also suggest that certain actions need clear incentives and broader organizational buy-in in order to drive adoption. 

\textit{Low resource, individual actions} do not require teams or significant resources and include considering data privacy implications, asking about the data or model to understand issues, and building/using transparency approaches. Among these actions, two did not show any correlation with organizational incentives and all three tended to have lower odds ratios (though still positive and statistically significant) with the other organizational conditions. These findings demonstrate that these actions are less dependent on organizational conditions and provide more space for individuals to exercise agency in adoption.

\section{Discussion}
\label{sec:discussion}
 In this discussion we first explore barriers and enablers to using genAI responsibly, before distinguishing between the two types of responsible AI practices that contribute to recoupling. Critically, and in contrast to past findings, we find that product managers view themselves as "guardrails" as opposed to "gatekeepers" for responsible AI use. This has implications for how PMs are conceptualized as responsible actors within the field of AI ethics and governance.

\subsection{Barriers to using generative AI responsibly}
There are three primary barriers to using genAI responsibly that product managers cited: uncertainty, diffused responsibility, and lack of incentives. Uncertainty existed at the level of the individual, the organization, and the industry, and contributed significantly to all of the other. Critically, too, it was not limited only to organizations but was endemic to the broader tech industry. This theme of uncertainty spillovers between individual, organization, and industry was a consistent theme across our findings. 

Uncertainty was exacerbated by lack of transparency about foundation models that product managers are using or integrating into products. Many product managers reported uncertainty about the inner workings of foundation models---including their training data, biases or potential risks---as this information is often unavailable, particularly if a product manager is leveraging a third-party model. This opacity is tied to industry-wide trends, where it remains common for corporate models to be closed, with organizations failing to disclose information about the training data and inner workings of the model. There is a concern around adopting models given that underlying risks in the model may not be publicly available or known to the product manager. Prior research has demonstrated that the black box nature of machine learning exacerbates and enforces lack of transparency in decision-making \cite{burrell2016how}. Our research expands those findings and shows how product managers using and developing AI tools struggle to justify their decision-making  when they  lack the appropriate tools to interpret the outputs of an AI system.

Diffused responsibility was closely tied with uncertainty at the organizational and industry level. Shared accountability coupled with lack of clear definitions and policies frequently led to inaction across teams. A common issue in organizations is the assumption by product managers that ethical considerations are being addressed by dedicated ethics or compliance units. Product managers assume that responsible actions are taken care of, especially if leadership does not emphasize shared responsibility across teams. Prior studies show that product managers were frequently seen as gatekeepers in ethical decision-making by other members of teams, such as engineers or designers \cite{rakova2021where, widder2023about, ali2023walking}. In our interviews, product managers discussed how they relied on other teams to bring ethical issues to their awareness but would often relegate decision-making to managers higher up, \textit{if} these concerns were even raised. This is discussed further in section 5.3.

Lack of incentives and diffused responsibility both contributed to moral hazard within organizations, when individuals neglect ethical priorities because of the sense that someone else is handling it \cite{holmstrom1979moral}. A key tension here (that was referenced numerous times in interviews) was the hype cycle in the tech industry. Product managers faced pressure to focus on speed to market and efficiency in shipping products that was  at odds with slowing down to consider ethical implications. This misalignment reflects a divergence between the goals of organizational leaders and the practices of product managers, due to the absence of incentives and accountability structures that encourage product managers to prioritize responsibility even when it means sacrificing some speed. The disconnect between stated principles and actual day-to-day practices and priorities highlights the occurrence of decoupling when AI principles are symbolically adopted but not substantially integrated into day-to-day workflows.  

\subsection{Enablers to using generative AI responsibly}
There are three primary enablers to using genAI responsibly that product managers cited: principles and policies, leadership commitment, and organizational incentives. Each of these can be seen as directly addressing the high levels of uncertainty faced by individuals, organizations, and the industry as a whole by providing greater transparency on how decisions are made. Principles, or high level policies and values, were frequently cited as playing a critical role in helping product managers navigate uncertainty and ambiguity. Critically, though, principles were not sufficient in and of themselves but needed to be coupled with either leadership commitment or resources. Product managers expressed a strong desire for greater education, tools, and processes to support responsible action as well as leadership commitment that would make them more confident when they raised ethical considerations. Absent this level of support, principles could exacerbate the problem of decoupling by providing a set of abstract policies or values that were challenging for product managers to interpret and translate on their own.

In addition to principles and leadership commitment, incentives were seen as critical to recognizing and rewarding individual behavior that enabled responsibility. Within the decoupling literature, organizations typically needed to experience either (1) external pressures that hold actors accountable to outside stakeholders, such as regulation, audits, or public scrutiny \cite{espeland1998struggle, hallett2010myth, turco2012difficult} or (2) commitment from higher management \cite{weaver1999integrated}. The way that external pressures and leadership commitment manifest within organizations is through incentive structures that align goals and demonstrate the prioritization of ethics. As we discuss in the following section, incentives played a critical role in promoting certain responsible AI practices that would not have been possible otherwise.

\subsection{Two types of responsible AI practices}
Responsible AI practices are not a monolith but rather a diverse set of individual and collective behaviors that can be adopted under varying conditions. One of the key contributions of this paper is in drawing a distinction between the different types of responsible AI practices that are commonly adopted and examining the organizational conditions that make them possible. We reviewed eight of the most common responsible genAI practices (see Table 2 in Section \ref{sec:findings}) to determine which four of the organizational conditions we have discussed (policies, principles, leadership commitment, incentives) support adoption among product managers. 

We find that leadership commitment, principles, and policies are positively associated with the adoption of all eight of the responsible practices we consider. Incentives, however, were not positively associated with two of the actions: considering data privacy implications and building or using transparency approaches. These findings highlight the universal importance of conditions such as leadership commitment, principles, and policies but also suggest that some actions require more activation energy than others. We break these eight practices into two categories: high resource, collective actions versus low resource, individual actions. 

\textit{High resource, collective actions} include: taking ethical / responsible AI training, using explainability methods, conducting fairness / bias testing, conducting audits of gen AI tools, and conducting adversarial testing or red teaming. What primarily distinguishes these actions is the high amount of resources they require, either in terms of infrastructure, labor, or time and/or the high amount of collective support that is necessary to put them into practice. For instance, taking ethical or responsible AI training requires organizations paying for access for courses for their employees or setting up these courses themselves. Practices such as conducting fairness/bias testing, audits, or adversarial testing require specialized teams, training, and tools that take time to set up. Due to the high start-up costs, these practices usually need to be approved at a higher level and require a significant amount of organizational buy-in across different teams. For this reason, they tend to be collective in nature. 

\textit{Low resource, individual actions }include: considering data privacy implications, building or using transparency approaches, and asking about the data or model to understand potential issues. These actions typically require far fewer resources in terms of infrastructure, labor, or time. For instance, considering data privacy implications can be as simple as an individual product manager deciding to read through the terms of service for a foundation model (or, more likely, entering the terms of service into a genAI tool to summarize). Due to lower start-up costs, these practices do not need to be approved at a higher level or have buy-in across the organization. They can be exercised by an individual at their own discretion. This is not to say that such actions do not carry risks---or that they are always carried out individually. Low resource, individual actions still show a strong positive association with policies, principles, and leadership commitment, which suggests that these practices do not come about organically. In interviews, respondents pointed to existing organizational cultures that prioritized values like protecting customer privacy or referenced existing training or company policies that encouraged checkpoints, reviews, and other safeguards to ensure high quality work. PMs, however, may be uniquely positioned to carry out some of these practices because of their access and oversight of multiple aspects of a particular product. Other roles may be more limited in the practices available to them.

These actions tended not to impede product development, yet they bridged the gap between formal principles and practical implementation, representing a form of recoupling. Even in an industry facing systemic uncertainty and in organizations with limited incentives, individual product managers found ways to take small-scale actions – such as protecting sensitive data and sharing information on AI systems---that gave them agency over the responsible use of genAI in their role. These actions also helped reduce diffusion of responsibility by modeling accountability and reinforcing a sense of responsibility in their teams. 

On a final note, certain practices exhibit very different size effects across the different organizational conditions. One hypothesis for the different effect sizes is that certain conditions and practices may be more likely to be included as part of a single organizational strategy. For example, organizations that release a set of organizational principles around responsible AI may also be likely to roll out responsible AI trainings for their employees (increased likelihood of 7.06 relative to baseline). On the other hand, instituting adversarial testing may be more likely to accompany the release of formal responsible AI policies by an organization (increased likelihood of 14.44 relative to baseline). We cannot confirm these hypotheses using our data but we see this as a promising avenue for future research.

\subsection{PMs as guardrails, not gatekeepers}
Interestingly, while prior literature portrayed product managers as gatekeepers to ethical decision-making within organizations, our study reveals that product managers are more likely to see themselves as guardrails, maintaining existing standards set up through organizational policies, principles, and direction from higher management \cite{rakova2021where, widder2023about, ali2023walking}. The fact that certain high-resource, collective actions were not available to product managers without broader organizational incentives suggests that their power is circumscribed. Especially in larger organizations, product managers typically saw the oversight of ethics as a function belonging to other, more specialized roles. Additionally, they viewed higher level management as the critical nexus point where ethical decision-making was made. 

These findings provide greater insight into how responsibility for ethical decision-making is viewed within cross-functional teams and within organizations more broadly. Decision-making on technologies is made within a social context. To properly understand how decisions are made it is necessary to understand how power is structured and leveraged within organizations. One of the central challenges of operationalizing AI ethics and governance is the lack of agency that individual actors feel within bureaucracies. Challenging the status quo requires taking on high degrees of individual risk, potentially jeopardizing one’s career or reputation \cite{morley2023operationalising}. The more risk individuals take on, the greater the potential for moral hazard, i.e. the incentive to push responsibility for ethical decision-making to others perceived to have more power or authority \cite{holmstrom1979moral}. As a result, we find that certain kinds of actions need collective support, often in the way of resources and broad institutional buy-in. 

This has implications for the role of product managers in supporting responsible use of genAI. While product managers have been viewed as gatekeepers by other members of their teams, such as engineers and designers, findings in the literature may reflect a bigger moral hazard that is endemic to the tech industry. To put it simply: if engineers and designers believe that product managers decide ethics and product managers believe that their higher level managers decide ethics, where does the chain end? Who ultimately bears the responsibility for ethical use of genAI? The challenge that the tech industry faces then is not so much in determining which single individual, role, or team is held responsible but in fostering an environment of collective responsibility and combating moral hazard.

\section{Conclusion}
\label{sec:conclusion}
This paper examined how product managers navigate the gap between ethical commitments and practice in genAI use for everyday work and product development. Drawing on interviews and a survey, we show that product managers operate under conditions of uncertainty, diffused responsibility, and misaligned incentives. Our findings challenge portrayals of product managers as ethical “gatekeepers”, who hold the key to making critical decisions in the product development cycle, showing instead that many function as maintainers of guardrails, upholding existing standards, principles, and policies rather than being primary arbiters of ethics. This reframing helps explain how ethical commitments become decoupled from practice despite apparent managerial authority. We further identify two pathways through which ethical commitments are partially enacted: low resource, individual actions that product managers can take autonomously, and high resource, collective actions that depend on organizational incentives and leadership support. We show that leadership commitment, principles, and policies are broadly associated with most responsible AI practices, but incentives are critical for enabling higher resource, collective forms of ethical action, such as auditing, adversarial testing, and fairness and bias testing. While individual actions provide a meaningful form of partial recoupling in otherwise constraining environments, relying on individual ethical agency as the primary mechanism to operationalize responsibility is limited and results in fragmentation. 

These findings carry implications for both organizational governance and AI regulation. Within organizations, accountability frameworks that assign responsibility to product teams without also mandating clear role definitions and aligned incentive structures are likely to replicate, rather than resolve, issues of uncertainty and diffused accountability. External pressures, including regulatory transparency requirements and industry standards, may be necessary to compel organizational incentive structures that support responsible AI action and enable consistency across product teams and organizations. 

There are several limitations in this study. First, interview and survey responses were self-reported. We did not observe respondents on the job. However, this also allowed us to get a sample across multiple organizations. Second, for the interviews, we used snowball sampling as a recruitment method, which may have led to respondents who already had a heightened interest and awareness of responsible AI. Third, all interviewees and most survey respondents are based in the US and clustered in tech or tech-adjacent industries. Future research on responsible use of genAI should be expanded to elucidate differences within and across different industries, as well as geographies. Research could also dive deeper within a small number of organizations and incorporate observational data, such as ethnography, so as not to rely on self-reported data. Finally, our regression analysis uses a binary outcome variable, which can obscure the nuances that drive organizations to adopt certain ethical practices and not others, though these nuances are explored in the qualitative results. Future research could build on these findings by examining the social-psychological factors and organizational conditions that drive adoption of individual responsible AI practices, using more granular outcome measures and qualitative methods to capture the complexity of decision-making involved.


\section*{Generative AI Usage Statement}
Generative AI was used to summarize existing text in a more concise format, including parts of the abstract, to generate prospective titles for the article, and to format the bibliography.

\section*{Ethical Considerations Statement}
\label{sec:ethics}
This paper examines the role of product managers (PMs) in the ethical governance of generative AI. While we hope these findings contribute to more accountable AI use and management practices, we acknowledge that findings could be misread as reducing expectations for product-team and product-level accountability. We caution against this interpretation. Our results do not suggest that PMs should bear less responsibility for ethics, but rather that this responsibility should be embedded into their day-to-day work through structural conditions and aligned incentives that make ethical action at the product team level genuinely possible.  Successful integration of responsible AI is not achieved by asking individuals to absorb ethical risk for an entire organization, but rather by aligning accountability structures and incentives across the organization and industry within which those individuals work.

For practitioners, we note that the data from this study informed a separate industry playbook that outlined actionable best practices for product managers and organizational leaders with the intention of making the findings of our research accessible to a broader audience. We encourage them to consult this resource for translatable research. \footnote{The playbook can be accessed here: \url{https://re-ai.berkeley.edu/sites/default/files/responsible_use_of_generative_ai_uc_berkeley_2025.pdf.}}

\begin{acks}
The authors are grateful to the interview and survey participants who generously shared their insights and made this research possible. We also thank the anonymous FAccT reviewers, whose feedback strengthened this paper, as well as the audiences at the Academy of Management and the American Sociological Association for their thoughtful comments. This research was supported by funding from Google. The views and findings expressed here are those of the authors and do not necessarily reflect those of Google.
\end{acks}

\bibliography{acmart}

\begin{appendices}


\clearpage

\section*{Research Instruments}

\subsection*{INTERVIEW GUIDE: Responsible Use of Generative AI in Organizations}
\label{app:interview_guide}

\subsection*{Background \& Perception of Incorporating Responsibility \& Ethics into Decisions}

\begin{enumerate}[leftmargin=*, label=\textbf{\arabic*.}]

  \item To start, can you tell me about your role and what brought you to where you are today?

  \item Does your team develop products that use gen AI models? (E.g., integrating a gen AI model into a new product)

        \begin{enumerate}[label=\textit{If Y\ldots}, leftmargin=2em]
          \item[] \begin{enumerate}[label=\alph*., leftmargin=1.5em]
            \item Can you provide examples?
            \item Who are the users? (internal vs.\ external; B2B vs.\ B2C)
            \item Do you use off-the-shelf gen AI models for the products?
            \item Do you optimize the gen AI models used, such as through fine-tuning, prompt engineering, or anything else?
          \end{enumerate}
        \end{enumerate}

  \item In regards to using gen AI for business purposes (e.g., marketing, data analysis): What kinds of gen AI tool(s) do you use and for what use cases?
        \textit{Probe:} For how long has it been used for different purposes?\
  \item What is the extent to which gen AI is used? For example, do most members of your team use gen AI tools frequently, or is it just used by some folks occasionally?

  \item What do you think it means or looks like to use gen AI models responsibly (whether in new products or in business use cases)?

        \textit{Note and probe for the following considerations:}
        \begin{itemize}[leftmargin=2em, label={--}]
          \item Fairness implications or potential bias
          \item Privacy and security concerns
          \item Hallucinations and misinformation
          \item Toxicity, harmful, and illegal content
          \item Environmental impacts
          \item How it impacts people's jobs and employment
        \end{itemize}
        \textit{Probe:} Do you prioritize certain aspects of responsible AI over others?

  \item Does your team take actions to use gen AI responsibly (whether in new products and/or in business operations)? If yes, what kinds of actions?

        \textit{Examples:} Work with fairness team prior to launching product, conduct audits, fairness/bias testing, adversarial testing, red teaming, building in transparency approaches, risk and impact assessments, and documentation.

        \textit{Note and probe for the following types of actions \emph{(Stahl et al., 2022)}:}
        \begin{itemize}[leftmargin=2em, label={--}]
          \item Technical approaches (e.g., bias testing, audits)
          \item Human oversight
          \item Ethical training and resource allocation for skill development
          \item Balancing competing goals and goods (e.g., slows down if needed for ethical checks)
        \end{itemize}

        \textit{Probe:} Have you/your team utilized any trainings, tools, or other support to more responsibly develop products that use gen AI? If so, what types and what has worked well in your experience?\

  \item What actions/resources have been most beneficial or helpful, and in what ways were they beneficial?

        \textit{Probe:} What actions/resources were least helpful?

  \item What were the motivations for you and your team to take the actions you shared before? Why do you do them?

  \item Do you think it is important for organizations to use gen AI responsibly --- either in using gen AI for new products or using gen AI for business operations? Why or why not?

        \textit{Probe:} What are the business benefits for responsible use and development of gen AI products?\\
        \textit{Probe:} What personal incentives do you see towards using gen AI responsibly?

\end{enumerate}

\subsection*{Challenges \& Tradeoffs in Using Gen AI Responsibly}

\begin{enumerate}[leftmargin=*, label=\textbf{\arabic*.}, resume]

  \item What are the challenges faced in using gen AI products responsibly (whether in new products and/or in business operations) and what lessons have you learned?

        \textit{Probe:} What are challenges faced by you personally? Your team?\\
        \textit{Probe:} What are challenges faced by the organization more broadly?

  \item Are there certain opportunity costs or tradeoffs that you think managers face in regards to using gen AI products responsibly?

        \textit{Probe:} How about opportunity costs for the business more broadly?\\
        \textit{Probe:} How do business pressures, such as speed to market, impact efforts to develop gen AI responsibly?\

\end{enumerate}

\subsection*{Conditions That Support Responsible Usage of Gen AI, Good Practices, \& Gaps}

\begin{enumerate}[leftmargin=*, label=\textbf{\arabic*.}, resume]

  \item Has your organization developed, adopted, or endorsed any principles for ethical or responsible AI? If so, are you familiar with them and what role do you think these principles play, if any, in supporting more responsible use of gen AI in your organization?

        \textit{Probe:} Have these principles influenced you or any of your colleagues, and if so, in what ways?\\
        \textit{Probe:} In implementing different responsible AI principles, have you ever needed to make choices in prioritizing some over others? How do you navigate these tradeoffs?

  \item Who is responsible in the organization for setting responsible AI approaches and requirements? Who is expected to put them into practice?

        \textit{Probe:} To what extent is ensuring responsibility when it comes to using gen AI part of your job duties or expectations?

  \item To what extent do you feel comfortable raising and talking about issues or concerns related to AI fairness or safety with colleagues?

        \textit{Probe:} How about with supervisors?\\
        \textit{Probe:} Can you share an example of a time where you did raise an issue or concern, and what happened afterwards?\
       
  \item What conditions do you think support more responsible usage of generative AI --- whether in products or within your team?

        \textit{Probe:} How do these conditions differ for different teams?\\
        \textit{Probe:} What is the role of leaders in creating enabling environments for responsible gen AI usage?\\
        \textit{Probe:} What is the role of tools/resources and education?

  \item In thinking about responsible use of gen AI, how do you engage with external actors?

        \textit{For example:} Examine what competitors are doing, engage with academics or other partners, participate in external working groups.

  \item What other resources, tools, or support would you like, if any, in regards to responsible use of gen AI?

        \textit{Probe:} What incentives should be adjusted or added to promote more responsible gen AI usage?

\end{enumerate}

\clearpage
\subsection*{SURVEY: Responsible Use of Generative AI Technologies in Organizations}
\label{app:survey}

\subsection*{Part 1. Demographic \& Background Information}

\begin{enumerate}[leftmargin=*, label=\textbf{D\arabic*.}]

  \item What country do you currently live in?

  \item Please select your age range.
        \begin{itemize}[leftmargin=2em, label=$\square$]
          \item 18--24 years old
          \item 25--34 years old
          \item 35--44 years old
          \item 45--54 years old
          \item 55--64 years old
          \item 65 years old or above
        \end{itemize}

  \item Please specify your race and ethnicity. \textit{Select all options that apply to you.}
        \begin{itemize}[leftmargin=2em, label=$\square$]
          \item American Indian, Alaska Native, or other Indigenous culture
          \item Black or African American
          \item East Asian
          \item Latino or Hispanic
          \item Native Hawaiian or Pacific Islander
          \item South Asian
          \item White
          \item Other:
        \end{itemize}

  \item What is your gender identity?
        \begin{itemize}[leftmargin=2em, label=$\square$]
          \item Female
          \item Male
          \item Non-binary
          \item Other: 
        \end{itemize}

  \item If you would like to, please elaborate on any of the above information.

\end{enumerate}

\begin{enumerate}[leftmargin=*, label=\textbf{OD\arabic*.}]

  \item What best describes your role?
        \begin{itemize}[leftmargin=2em, label=$\square$]
          \item Product Manager / Lead
                \begin{itemize}[leftmargin=2em, label=$\circ$]
                  \item Technical Product Manager
                  \item Data/analytics Product Manager
                  \item UX Product Manager
                  \item Product Lead
                  \item Other type of Product Manager/Lead: 
                \end{itemize}
          \item Engineer / data scientist
          \item Product marketing / business development
          \item Product designer / UX / content designer
          \item Other: 
        \end{itemize}

  \item What is the approximate number of employees at your company (total globally)?
        \begin{itemize}[leftmargin=2em, label=$\square$]
          \item 1--49
          \item 50--499
          \item 499--1,000
          \item 1,000--4,999
          \item 5,000 or more
          \item Don't know
          \item Other
        \end{itemize}

  \item What industry best describes the nature of your company's business?
        \begin{itemize}[leftmargin=2em, label=$\square$]
          \item Agriculture and Farming
          \item Arts and Entertainment
          \item Automotive
          \item Construction and Real Estate
          \item Customer Relationship Management
          \item Education
          \item Finance and Insurance
          \item Healthcare and Pharmaceuticals
          \item Hospitality and Tourism
          \item Information Technology (IT) and Software
          \item Manufacturing
          \item Retail
          \item Telecommunications
          \item Transportation and Logistics
          \item Other (please specify):
        \end{itemize}

  \item Where is the company headquarters? 

\end{enumerate}

\subsection*{Part 2. Survey Questions}
 
\begin{enumerate}[leftmargin=*, label=\textbf{Q\arabic*.}]
 
  \item For what types of work tasks/purposes do you use gen AI? \textit{[Check all that apply]}
        \begin{itemize}[leftmargin=2em, label=$\square$]
          \item Content generation (e.g., written content to support marketing and customer engagement)
          \item Productivity tasks (e.g., summarizing meeting notes, writing emails)
          \item UI/UX design
          \item Product ideation and brainstorming
          \item Data analysis and insights
          \item Coding
          \item Customer support automation
          \item As a component of a product we are developing (e.g., using/integrating gen AI in new products)
          \item Other: 
        \end{itemize}
 
  \item Why are you using gen AI (or not) for certain work tasks/purposes? \textit{[Open ended]}
 
  \item What gen AI tools/models do you use? \textit{[Check all that apply]}
        \begin{itemize}[leftmargin=2em, label=$\square$]
          \item My company's own gen AI tools/models developed by our research teams using our own foundation model(s)\\
                \textit{Please share more about the tool(s)/model(s):} \underline{\hspace{3cm}}
          \item My company's own gen AI tools/models developed by our research teams based on third-party foundation model(s)\\
                \textit{Please share more about the tool(s)/model(s):} 
          \item External proprietary gen AI tools/models (e.g., ChatGPT)\\
                \textit{If you use external tool(s)/model(s), please enter their names here:} 
          \item External open-source gen AI tools/models (e.g., LLaMA)\\
                \textit{If you use external tool(s)/model(s), please enter their names here:}
        \end{itemize}
 
  \item In using gen AI, do you optimize models, such as through fine tuning models or prompt engineering? Please explain.
 
  \item What do you think it means or looks like to use gen AI tools responsibly? \textit{[Open ended]}
 
  \item When considering whether and how to use gen AI within your team, do you and your team consider how to use gen AI tools responsibly? Why or why not? \textit{[Open ended]}
 
  \item How could responsible use of gen AI impact business outcomes (positively and/or negatively)? \textit{[Open ended]}
 
  \item To what extent do you agree or disagree with the following statement: \textit{``Being responsible in using gen AI is important to gain business benefits.''}
 
        \begin{itemize}[leftmargin=2em, label=$\bigcirc$]
          \item Strongly disagree
          \item Disagree
          \item Neither agree nor disagree
          \item Agree
          \item Strongly agree
        \end{itemize}
 
  \item Do you take any of the following actions when using gen AI tools? \textit{Check all that apply.}
        \begin{itemize}[leftmargin=2em, label=$\square$]
          \item Work with colleagues whose job responsibilities are related to supporting AI fairness or responsibility
          \item Take ethical/responsible AI trainings
          \item Conduct audits of the gen AI tools
          \item Conduct fairness/bias testing
          \item Consider data privacy implications and take actions to protect data privacy
          \item Ask about the data or model to understand potential limitations or issues it has
          \item Conduct adversarial testing or red teaming
          \item Use explainability methods (e.g., to enable team members to better understand and probe the model)
          \item Build transparency approaches (e.g., use documentation that can make dataset and model decisions transparent to others)
          \item No actions taken (to my knowledge)
          \item Other: \underline{\hspace{4cm}}
        \end{itemize}
 
        \noindent\textbf{9.1} Please build on any of the actions checked above:
 
        \noindent\textbf{9.2} How do any of the actions you selected above depend or differ based on the particular gen AI use case?
 
  \item Do you have any formal training in using gen AI responsibly?
        \begin{itemize}[leftmargin=2em, label=$\bigcirc$]
          \item Yes
          \item No
        \end{itemize}
 
        \noindent\textit{If yes:} What is that training, and when did you receive it (i.e., before your current job, in college, within your current job)?
 
  \item Does your organization have\ldots{} \textit{[Check all that apply]}
        \begin{itemize}[leftmargin=2em, label=$\square$]
          \item Leadership that has expressed commitment to responsible AI
          \item Responsible or ethical AI principles
          \item A policy/policies that inform the use of gen AI
          \item Clear incentives for using/implementing gen AI responsibly
          \item None
          \item I don't know
        \end{itemize}
 
        \noindent\textbf{11.1} \textit{If your organization has AI principles or a policy on use of gen AI:} How does that influence your use of gen AI (or not)?
 
        \noindent\textbf{11.2} \textit{If there are clear incentives:} Can you please share more about what these incentives are?
 
  \item What types of challenges have you or your team faced in regards to using gen AI responsibly? \textit{[Check all that apply]}
        \begin{itemize}[leftmargin=2em, label=$\square$]
          \item Lack of clarity on what that looks like
          \item Lack of training
          \item Lack of resources or tools
          \item Lack of incentives
          \item Lack of understanding whether/why it may be valuable
          \item Lack of support
          \item Lack of clarity about expectations or relevance to my role
          \item None
          \item Other: \underline{\hspace{4cm}}
        \end{itemize}
 
  \item On a scale of 1 to 5, to what extent do you agree that the following are challenges to using AI responsibly?
 
        \textit{Scale: 1 = Strongly disagree \quad 2 = Disagree \quad 3 = Neutral \quad 4 = Agree \quad 5 = Strongly agree}
 
        \begin{tabular}{lcc}
          \toprule
          \textbf{Statement} & \textbf{Rating (1--5)} \\
          \midrule
          Business pressures (e.g., speed to market) & \underline{\hspace{2cm}} \\[4pt]
          Concerns about being labeled a ``troublemaker'' & \underline{\hspace{2cm}} \\[4pt]
          Discomfort raising issues around responsibility & \underline{\hspace{2cm}} \\[4pt]
          Other: \underline{\hspace{5cm}} & \underline{\hspace{2cm}} \\
          \bottomrule
        \end{tabular}
 
  \item Are there any tradeoffs or opportunity costs in implementing responsible AI approaches? If yes, what are these and how do you navigate them? \textit{[Open ended]}
 
  \item To what extent do you agree with the following statements?
 
        \textit{Scale: 1 = Strongly disagree \quad 2 = Disagree \quad 3 = Neutral \quad 4 = Agree \quad 5 = Strongly agree \quad N/A}
 
        \begin{tabular}{lc}
          \toprule
          \textbf{Statement} & \textbf{Rating} \\
          \midrule
          I have suggestions for my supervisor about how to use gen AI more responsibly. & \underline{\hspace{2cm}} \\[6pt]
          I feel comfortable volunteering suggestions to my supervisor about how to & \\
          \quad use gen AI more responsibly. & \underline{\hspace{2cm}} \\[6pt]
          My supervisor asks and/or encourages me to think about and take actions & \\
          \quad to help the team be more responsible in using gen AI. & \underline{\hspace{2cm}} \\
          \bottomrule
        \end{tabular}
 
  \item What conditions in your team or organization do you think support more responsible usage of gen AI? \textit{[Open ended]}
 
  \item What other resources, tools, or support would you like, if any, in regards to responsible use of gen AI? \textit{[Open ended]}
 
  \item Is there any other information you would like to add or share? \textit{[Open ended]}
 
  \item Would you be open to a short follow-up interview exploring these topics?
        \begin{itemize}[leftmargin=2em, label=$\bigcirc$]
          \item Yes
          \item No
        \end{itemize}
 
        \textit{If yes, you will be directed to the interest form.}
 
\end{enumerate}
 
\noindent\textbf{Thank you for completing the survey!} To protect your privacy, it is recommended to clear your browser's history, cache, cookies, and other browsing data.

\end{appendices}

\end{document}